\input harvmac

\noblackbox

\let\includefigures=\iftrue
\let\useblackboard=\iftrue
\newfam\black

\includefigures
\message{If you do not have epsf.tex (to include figures),}
\message{change the option at the top of the tex file.}
\input epsf
\def\figin{\epsfcheck\figin}\def\figins{\epsfcheck\figins}
\def\epsfcheck{\ifx\epsfbox\UnDeFiNeD
\message{(NO epsf.tex, FIGURES WILL BE IGNORED)}
\gdef\figin##1{\vskip2in}\gdef\figins##1{\hskip.5in}
\else\message{(FIGURES WILL BE INCLUDED)}%
\gdef\figin##1{##1}\gdef\figins##1{##1}\fi}
\def\DefWarn#1{}
\def\figinsert{\goodbreak\midinsert}
\def\ifig#1#2#3{\DefWarn#1\xdef#1{Fig.~\the\figno}
\writedef{#1\leftbracket Fig.\noexpand~\the\figno}%
\figinsert\figin{\centerline{#3}}\medskip\centerline{\vbox{
\baselineskip12pt\advance\hsize by -1truein
\noindent\footnotefont{\bf Fig.~\the\figno:} #2}}
\bigskip\endinsert\global\advance\figno by1}
\else
\def\ifig#1#2#3{\xdef#1{Fig.~\the\figno}
\writedef{#1\leftbracket Fig.\noexpand~\the\figno}%
\global\advance\figno by1} \fi

\def\doublefig#1#2#3#4{\DefWarn#1\xdef#1{Fig.~\the\figno}
\writedef{#1\leftbracket Fig.\noexpand~\the\figno}%
\figinsert\figin{\centerline{#3\hskip1.0cm#4}}\medskip\centerline{\vbox{
\baselineskip12pt\advance\hsize by -1truein
\noindent\footnotefont{\bf Fig.~\the\figno:} #2}}
\bigskip\endinsert\global\advance\figno by1}

\useblackboard
\message{If you do not have msbm (blackboard bold) fonts,}
\message{change the option at the top of the tex file.}
\font\blackboard=msbm10 scaled \magstep1 \font\blackboards=msbm7
\font\blackboardss=msbm5 \textfont\black=\blackboard
\scriptfont\black=\blackboards \scriptscriptfont\black=\blackboardss

\else

\fi
%
\def\subsubsec#1{\bigskip\noindent{\it{#1}} \bigskip}
\def\yboxit#1#2{\vbox{\hrule height #1 \hbox{\vrule width #1
\vbox{#2}\vrule width #1 }\hrule height #1 }}
\def\fillbox#1{\hbox to #1{\vbox to #1{\vfil}\hfil}}
\def\ybox{{\lower 1.3pt \yboxit{0.4pt}{\fillbox{8pt}}\hskip-0.2pt}}
%
%


\def\comments#1{}



\def\II{\relax{I\kern-.10em I}}

\def\IZ{\relax{\rm Z\kern-.34em Z}}
\def\IB{\relax{\rm I\kern-.18em B}}
\def\IC{{\relax\hbox{$\inbar\kern-.3em{\rm C}$}}}
\def\ID{\relax{\rm I\kern-.18em D}}
\def\IE{\relax{\rm I\kern-.18em E}}
\def\IF{\relax{\rm I\kern-.18em F}}
\def\IG{\relax\hbox{$\inbar\kern-.3em{\rm G}$}}
\def\IGa{\relax\hbox{${\rm I}\kern-.18em\Gamma$}}
\def\IH{\relax{\rm I\kern-.18em H}}
\def\II{\relax{\rm I\kern-.18em I}}
\def\IK{\relax{\rm I\kern-.18em K}}
\def\IP{\relax{\rm I\kern-.18em P}}

%

\def\inbar{\,\vrule height1.5ex width.4pt depth0pt}

\def\IR{\relax{\rm I\kern-.18em R}}

\def\simgt{\hskip0.05in\relax{
\raise3.0pt\hbox{ $>$ {\lower5.0pt\hbox{\kern-1.05em $\sim$}} }}
\hskip0.05in}

%


%

\def\lp10{\ell_p^{10}}
\def\lp11{\ell_p^{11}}
\def\R11{R_{11}}

\def\frac#1#2{{#1 \over #2}}



\newdimen\tableauside\tableauside=1.0ex
\newdimen\tableaurule\tableaurule=0.4pt
\newdimen\tableaustep
\def\phantomhrule#1{\hbox{\vbox to0pt{\hrule height\tableaurule width#1\vss}}}
\def\phantomvrule#1{\vbox{\hbox to0pt{\vrule width\tableaurule height#1\hss}}}
\def\sqr{\vbox{%
  \phantomhrule\tableaustep
  \hbox{\phantomvrule\tableaustep\kern\tableaustep\phantomvrule\tableaustep}%
  \hbox{\vbox{\phantomhrule\tableauside}\kern-\tableaurule}}}
\def\squares#1{\hbox{\count0=#1\noindent\loop\sqr
  \advance\count0 by-1 \ifnum\count0>0\repeat}}
\def\tableau#1{\vcenter{\offinterlineskip
  \tableaustep=\tableauside\advance\tableaustep by-\tableaurule
  \kern\normallineskip\hbox
    {\kern\normallineskip\vbox
      {\gettableau#1 0 }%
     \kern\normallineskip\kern\tableaurule}%
  \kern\normallineskip\kern\tableaurule}}
\def\gettableau#1 {\ifnum#1=0\let\next=\null\else
  \squares{#1}\let\next=\gettableau\fi\next}

\tableauside=1.0ex \tableaurule=0.4pt


 %
 %
 \def\eqnn#1{\xdef #1{(\secsym\the\meqno)}\writedef{#1\leftbracket#1}%
 \global\advance\meqno by1\wrlabeL#1}
 \def\eqna#1{\xdef #1##1{\hbox{$(\secsym\the\meqno##1)$}}
 \writedef{#1\numbersign1\leftbracket#1{\numbersign1}}%
 \global\advance\meqno by1\wrlabeL{#1$\{\}$}}
 \def\eqn#1#2{\xdef #1{(\secsym\the\meqno)}\writedef{#1\leftbracket#1}%
 \global\advance\meqno by1$$#2\eqno#1\eqlabeL#1$$}

\global\newcount\itemno \global\itemno=0

\def\itemaut#1{\global\advance\itemno by1\noindent\item{\the\itemno.}#1}



\def\ie{{\it i.e.}}

\hyphenation{Di-men-sion-al}



\lref\CdL{
  S.~R.~Coleman and F.~De Luccia,
  ``Gravitational Effects On And Of Vacuum Decay,''
  Phys.\ Rev.\ D {\bf 21}, 3305 (1980).
}

\lref\tometal{A. Aguirre, T. Banks, M. Dine, M. Johnson, A. Shomer,
work in progress}

\lref\crosssection{
  W.~G.~Unruh,
  ``Absorption Cross-Section Of Small Black Holes,''
  Phys.\ Rev.\ D {\bf 14}, 3251 (1976).
}

\lref\Qball{N. Arkani-Hamed, P. Schuster, N. Toro, ``Observational
Constraints on the Vacuum of our Universe from Catalyzed Vacuum
Decay", to appear.}

\lref\graveyard{
  A.~Gould, B.~T.~Draine, R.~W.~Romani and S.~Nussinov,
  ``Neutron Stars: Graveyard Of Charged Dark Matter,''
  Phys.\ Lett.\ B {\bf 238}, 337 (1990).
}

\lref\trapping{
  L.~Kofman, A.~Linde, X.~Liu, A.~Maloney, L.~McAllister and E.~Silverstein,
  ``Beauty is attractive: Moduli trapping at enhanced symmetry points,''
  JHEP {\bf 0405}, 030 (2004)
  [arXiv:hep-th/0403001].
}

\lref\BHtrap{
  N.~Kaloper, J.~Rahmfeld and L.~Sorbo,
  ``Moduli entrapment with primordial black holes,''
  Phys.\ Lett.\ B {\bf 606}, 234 (2005)
  [arXiv:hep-th/0409226].
}

\lref\DamourPoly{
  T.~Damour and A.~M.~Polyakov,
  ``The String dilaton and a least coupling principle,''
  Nucl.\ Phys.\ B {\bf 423}, 532 (1994)
  [arXiv:hep-th/9401069].
}

\lref\KhouryAQ{
  J.~Khoury and A.~Weltman,
  ``Chameleon fields: Awaiting surprises for tests of gravity in space,''
  Phys.\ Rev.\ Lett.\  {\bf 93}, 171104 (2004)
  [arXiv:astro-ph/0309300].
}

\lref\garyjoe{
  G.~T.~Horowitz and J.~Polchinski,
  ``Instability of spacelike and null orbifold singularities,''
  Phys.\ Rev.\ D {\bf 66}, 103512 (2002)
  [arXiv:hep-th/0206228].
}

\lref\spacelike{
  J.~McGreevy and E.~Silverstein,
  ``The tachyon at the end of the universe,''
  JHEP {\bf 0508}, 090 (2005)
  [arXiv:hep-th/0506130].
}

\lref\finalstate{
  G.~T.~Horowitz and J.~Maldacena,
  ``The black hole final state,''
  JHEP {\bf 0402}, 008 (2004)
  [arXiv:hep-th/0310281];
  G.~T.~Horowitz and E.~Silverstein,
  ``The inside story: Quasilocal tachyons and black holes,''
  Phys.\ Rev.\ D {\bf 73}, 064016 (2006)
  [arXiv:hep-th/0601032].
}

\lref\smolin{
  L.~Smolin,
  ``Cosmological natural selection as the explanation for the complexity of the
  universe,''
  PhysicaA {\bf 340}, 705 (2004);
  E.~J.~Martinec,
  ``Space - like singularities and string theory,''
  Class.\ Quant.\ Grav.\  {\bf 12}, 941 (1995)
  [arXiv:hep-th/9412074].
}

\lref\BHmech{J. M. Bardeen, B. Carter, and S. Hawking, ``The Four
Laws of Black Hole Mechanics", Commun.\ Math.\ Phys.\  {\bf 31}, 161
(1973)}

\lref\page{
  S.~W.~Hawking,
  ``Black Hole Explosions,''
  Nature {\bf 248}, 30 (1974).
  S.~W.~Hawking,
  ``Particle Creation By Black Holes,''
  Commun.\ Math.\ Phys.\  {\bf 43}, 199 (1975)
  [Erratum-ibid.\  {\bf 46}, 206 (1976)].
  D.~N.~Page,
  ``Particle Emission Rates From A Black Hole. III. Charged Leptons From A
  Nonrotating Hole,''
  Phys.\ Rev.\ D {\bf 16}, 2402 (1977);
}

\lref\land{
  R.~Bousso and J.~Polchinski,
  ``Quantization of four-form fluxes and dynamical neutralization of the
  cosmological constant,''
  JHEP {\bf 0006}, 006 (2000)
  [arXiv:hep-th/0004134];
  S.~B.~Giddings, S.~Kachru and J.~Polchinski,
  ``Hierarchies from fluxes in string compactifications,''
  Phys.\ Rev.\ D {\bf 66}, 106006 (2002)
  [arXiv:hep-th/0105097];
  E.~Silverstein,
  ``(A)dS backgrounds from asymmetric orientifolds,''
  Contributed to Strings 2001,
  [arXiv:hep-th/0106209];
  A.~Maloney, E.~Silverstein and A.~Strominger,
  ``De Sitter space in noncritical string theory,''
Published in *Cambridge 2002, The future of theoretical physics and
cosmology* 570-591, [arXiv:hep-th/0205316];
  B.~S.~Acharya,
  ``A moduli fixing mechanism in M theory,''
  arXiv:hep-th/0212294.
  S.~Kachru, R.~Kallosh, A.~Linde and S.~P.~Trivedi,
  Phys.\ Rev.\ D {\bf 68}, 046005 (2003)
  [arXiv:hep-th/0301240].
{\it et seq.}}

\lref\andrei{
  A.~M.~Green and A.~R.~Liddle,
  ``Constraints on the density perturbation spectrum from primordial black
  holes,''
  Phys.\ Rev.\ D {\bf 56}, 6166 (1997)
  [arXiv:astro-ph/9704251].
  J.~Garcia-Bellido, A.~D.~Linde and D.~Wands,
  ``Density perturbations and black hole formation in hybrid inflation,''
  Phys.\ Rev.\ D {\bf 54}, 6040 (1996)
  [arXiv:astro-ph/9605094].
}

\lref\mavens{
  R.~Fardon, A.~E.~Nelson and N.~Weiner,
  ``Dark energy from mass varying neutrinos,''
  JCAP {\bf 0410}, 005 (2004)
  [arXiv:astro-ph/0309800].
}

\lref\raffelt{
  G.~G.~Raffelt,
  ``Stars as laboratories for fundamental physics: The astrophysics of
  neutrinos, axions, and other weakly interacting particles "
Chicago, USA: Univ. Pr. (1996) 664 p.
}

\lref\woosley{
  S.~Woosley and T.~Janka,
  ``The Physics of Core-Collapse Supernovae,''
  arXiv:astro-ph/0601261.
}

\lref\burrows{
  A.~Burrows, E.~Livne, L.~Dessart, C.~Ott and J.~Murphy,
  ``A New Mechanism for Core-Collapse Supernova Explosions,''
  arXiv:astro-ph/0510687.
}

\lref\sandipetal{
  K.~Goldstein, N.~Iizuka, R.~P.~Jena and S.~P.~Trivedi,
  ``Non-supersymmetric attractors,''
  Phys.\ Rev.\ D {\bf 72}, 124021 (2005)
  [arXiv:hep-th/0507096].
}

\lref\stringgas{
  T.~Battefeld and S.~Watson,
  ``String gas cosmology,''
  arXiv:hep-th/0510022.
}

\lref\DVB{ V.A. Berezin, V.A. Kuzmin and I.I. Tkachev, ``Dynamics of
Bubbles in General Relativity", Phys. Rev. D 36, 2919}

\lref\cattun{
  I.~K.~Affleck and F.~De Luccia,
  ``Induced Vacuum Decay,''
  Phys.\ Rev.\ D {\bf 20}, 3168 (1979).
  M.~B.~Voloshin,
  ``Catalyzed decay of false vacuum in four-dimensions,''
  Phys.\ Rev.\ D {\bf 49}, 2014 (1994).
  V.~A.~Rubakov and D.~T.~Son,
  ``Instanton like transitions at high-energies in (1+1)-dimensional scalar
  models. 2. Classically allowed induced vacuum decay,''
  Nucl.\ Phys.\ B {\bf 424}, 55 (1994)
  [arXiv:hep-ph/9401257].
  V.~A.~Berezin, V.~A.~Kuzmin and I.~I.~Tkachev,
  ``Black Holes Initiate False Vacuum Decay,''
  Phys.\ Rev.\ D {\bf 43}, 3112 (1991);
  A.~Gomberoff, M.~Henneaux, C.~Teitelboim and F.~Wilczek,
  ``Thermal decay of the cosmological constant into black holes,''
  Phys.\ Rev.\ D {\bf 69}, 083520 (2004)
  [arXiv:hep-th/0311011].
}

\lref\attractorsusy{
  S.~Ferrara, R.~Kallosh and A.~Strominger,
  ``N=2 extremal black holes,''
  Phys.\ Rev.\ D {\bf 52}, 5412 (1995)
  [arXiv:hep-th/9508072].
}

\lref\ulf{
  U.~H.~Danielsson, A.~Guijosa and M.~Kruczenski,
  ``Brane-antibrane systems at finite temperature and the entropy of black
  branes,''
  JHEP {\bf 0109}, 011 (2001)
  [arXiv:hep-th/0106201].
}

\lref\garybubbles{
  G.~T.~Horowitz,
  ``Tachyon condensation and black strings,''
  JHEP {\bf 0508}, 091 (2005)
  [arXiv:hep-th/0506166];
  S.~F.~Ross,
  ``Winding tachyons in asymptotically supersymmetric black strings,''
  JHEP {\bf 0510}, 112 (2005)
  [arXiv:hep-th/0509066].
}

\lref\unruh{
  W.~G.~Unruh and N.~Weiss,
  ``Acceleration radiation in interacting field theories,''
  Phys. Rev. D {\bf 29}, 1656 (1984);
}

\Title{\vbox{\baselineskip12pt\hbox{}
\hbox{SU-ITP-06/13}\hbox{SLAC-PUB-11846}}} {\vbox{
\centerline{Attractor Explosions and Catalyzed Vacuum Decay}
\smallskip
}}
\bigskip
\bigskip
\centerline{Daniel Green, Eva Silverstein, and David Starr}
\bigskip
\bigskip
\centerline{{\it SLAC and Department of Physics, Stanford University, Stanford, CA 94305-4060}}
\bigskip
\bigskip
\noindent

We present a mechanism for catalyzed vacuum bubble production
obtained by combining moduli stabilization with a generalized
attractor phenomenon in which moduli are sourced by compact objects.
This leads straightforwardly to a class of examples in which the
Hawking decay process for black holes unveils a bubble of a
different vacuum from the ambient one, generalizing the new endpoint
for Hawking evaporation discovered recently by Horowitz. Catalyzed
vacuum bubble production can occur for both charged and uncharged
bodies, including Schwarzschild black holes for which massive
particles produced in the Hawking process can trigger vacuum decay.
We briefly discuss applications of this process to the population
and stability of metastable vacua.


\bigskip
\Date{May 2006}

\newsec{Introduction}

Recently much progress has been made in stabilizing moduli and
analyzing the resulting discretuum of vacua \land.  Another topic of
recent interest has been the attractor mechanism (see e.g.
\refs{\attractorsusy,\sandipetal}), in which moduli are driven to
fixed values at the horizon of charged black holes.

In general, both types of effects are present.  The dynamics of moduli $\phi_i$ are determined by the scalar
potential ${\cal U}(\phi)$ and by other local sources $\rho (\phi;x)$ to which the moduli couple via an equation
of motion of the form
\eqn\generalcoup{\nabla^2\phi\sim -\del_\phi{\cal U}(\phi) -
\del_\phi \rho}
For example, in weakly coupled string theory, the dilaton $\Phi$
modulates masses and couplings of particles, so energy densities
formed from them typically depend on $\Phi$. As a specific example,
moduli-dependence in the kinetic terms for electromagnetic fields
yields forces on moduli near charged sources, an ingredient in the
attractor mechanism.

Taking both effects into account leads to the possibility of
catalyzed production of bubbles of different vacua. Consider the
subset of backgrounds where structure formation leads to objects
dense enough that the source terms $\rho$ in \generalcoup\ compete
with the barrier height in $\cal U$ separating one metastable
minimum of ${\cal U}$ from another. The moduli can then be locally
forced into the basin of attraction of a different minimum from that
at which they are metastabilized in bulk.  Depending on the
parameters, this can result in vacuum decay or in production of a
perturbatively stable nonexpanding bubble of a different vacuum.   A
dramatic string-theoretic example of this was obtained recently in
\garybubbles, where certain charged sources produce a tachyonic
instability nearby, catalyzing the formation of a bubble of nothing.

One goal of the present note is to point out a large class of
examples where bubbles of other vacua appear in the course of
Hawking evaporation of black holes.  In particular, in the charged
case we generalize \garybubbles\ and in the uncharged case we obtain
a new mechanism for vacuum bubble production.

For charged black holes, this arises in variants of the process
\garybubbles\ accessible in effective field theory. One can start in
one metastable vacuum and form a black hole with mass much greater
than its charge.  At first it behaves approximately like a
Schwarzschild black hole, gradually losing mass to Hawking
radiation. At some point long before it reaches the correspondence
point, the black hole has radiated away enough of its mass that the
charge plays a role, exerting local forces on the scalar fields in
the problem and creating a bubble of a different vacuum.

Vacuum bubbles can also be produced in the case of uncharged compact
objects, including Schwarzschild black holes, via the accumulation
of uncharged matter sourcing scalar fields. Depending on the
relative strength of the local energy density and the ambient moduli
potential, the bubbles may appear outside or inside the
Schwarzschild radius. Bubbles may form outside in two ways:  in the
process of collapse before the horizon forms, and in the process of
Hawking evaporation as the black hole becomes small enough to
produce massive Hawking particles sourcing the scalar.  In the
latter case, we will present a window of parameters for which this
new mechanism constitutes the dominant method for producing vacuum
bubbles, occurring faster than tunneling and other effects in the
black hole background.

One application is to black hole physics, whose dynamics contains
information about other vacua.  This plays a role in the
microphysical description of black holes and their information flow.

Another application is to provide a method for populating the
landscape which proceeds perturbatively. Backgrounds with larger
variations in density are more likely to produce vacuum bubbles;
conversely backgrounds with less dense structure formation are more
stable.\foot{There have been a number of calculations of catalyzed
tunneling decays in field theory and gravity \cattun. For simplicity
we will here focus on the regime where local overdensities develop
strongly enough to classically kick the scalar field over a
potential barrier separating different local minima.  There have
been a number of works using gases of particles obtained
cosmologically to provide transient forces on scalar fields, as in
thermal inflation and e.g.
\refs{\DamourPoly,\trapping,\BHtrap,\stringgas}.}
 The process also provides a dynamical mechanism favoring regions
 with light particles; this aspect is similar to  moduli trapping \trapping\ except
 here the particle production is effected by structure formation and evaporating black
 holes.

Perhaps the most compelling application of this process would be to
obtain regions of parameter space where catalyzed bubble production
would be possible realistically, probing regions of the landscape.
As we discuss, this depends on the degree to which dense structures
and small black holes form as a result of early universe inflation
and structure formation.  A much more extensive treatment of this
appears in the interesting related work of \Qball\ to appear. The
prospect of catalyzed decays and their realistic embedding is also
under investigation in \tometal.

The paper is organized as follows.  In section 2 we set up two
canonical systems where moduli are sourced by local densities,  and
review bubble dynamics in a way applicable to the catalyzed
formation process.  The main new results are contained in section 3,
where we present the catalyzed bubble production mechanisms for
charged and uncharged black holes.  Section 4 describes various
applications of the mechanism.

\newsec{Catalyzed Bubble production and evolution}

Consider for simplicity first a regime in which low energy field theory is a good approximation.  We will be
interested in the situation described above \generalcoup\ in which the scalar equation of motion can be written
in the form
\eqn\genscalar{\nabla^2\phi=-\del_\phi {\cal U}_{tot}(r, \phi)}
where the total effective potential ${\cal U}_{tot}$ depends on
radial coordinate distance $r$ from the source, in a metric of the
form
\eqn\genmet{ds^2=-a(r)^2dt^2+{{dr^2}\over{a(r)^2}}+b(r)^2 d\Omega^2}
in terms of functions $a(r),b(r)$ which approach $1$ and $r$
respectively in the limit where gravitational effects are
negligible.

We will find it useful to parameterize the moduli potential in terms
of its energy scale $M_U$ and the typical size of its features in
field space $M_0$:
\eqn\Uparam{{\cal U}=M_U^4 f(\phi/M_0)}
In particular, the barrier width is of order $M_0$.

Let us give two illustrative examples.  First, we can consider a charged compact object, with scalars $\phi_i$
coupling to electromagnetic fields according to the action
\eqn\EMac{S= \int d^4x \sqrt{-G} \biggl(M_p^2 R - 2 (\partial
\phi_i)^2-f_{ab}(\phi_i) F^a_{\mu \nu} F^{b{\mu\nu}}-{\cal
U}(\phi_i)\biggr)}
(generalizing that in \sandipetal\ to include the ambient moduli
potential ${\cal U}$).  For simplicity we will focus on a single
direction $\phi$ scalar field space.  Given spherical symmetry and
Gauss' law, we can write say for a magnetically charged black hole
\eqn\Fform{F^a=Q^a \sin\theta d\theta\wedge d\phi}
The equations of motion for the scalar field in this background are
\eqn\scaleq{{1\over {b^2\sin\theta}}{\partial_{\mu}(b^2 \sin\theta
\,
\partial^{\mu}\phi) = \biggl({{V_{eff}^{\prime}} \over {2b^4}} + \frac{1}{4}{\cal U}^{\prime}\biggr) (\phi)}}
where $^\prime$ denotes differentiation with respect to $\phi$ and
where the local contribution to the force on the scalar arises from
an effective potential
\eqn\Veff{V_{eff}=f_{ab}(\phi)Q^aQ^b}
obtained from the electromagnetic energy by plugging \Fform\ into
\EMac. The total effective potential in the sense of \genscalar\ is
${\cal U}_{tot}=V_{eff}/b^4+{\cal U}$.

As a second illustrative example, let us consider a scalar $\phi$ which modulates the mass of a particle, for
example a fermion field $\psi$.  The action governing this sector of the theory is
\eqn\Aac{S= \int d^4x \sqrt{-G}\biggl(M_p^2 R + i\bar\psi D\psi- 2
(\partial \phi)^2 + \tilde f(\phi/\tilde M_0)m_\psi
\bar\psi\psi-{\cal U}(\phi)+\Omega\bar\psi \gamma^0\psi\biggr)}
where we included the possibility of a chemical potential $\Omega$,
and we parameterize the argument of the modulating function $f$ via
a scale $\tilde M_0$. In a background density of $\psi$ particles,
the scalar field equation of motion can again be written in the form
\scaleq, where in this case the local effective potential is
\eqn\VeffII{{V_{eff}\over b^4}= \tilde f(\phi/\tilde M_0) \langle
m_\psi \bar\psi \psi \rangle (r,t)\equiv \rho_\psi(r,t;\phi)}

In a background which includes the evolution of structure, the
massive fields $\psi$ will develop inhomogeneities (along with other
sectors of the theory).  We will be interested in cases where these
inhomogeneities make the $\partial_\phi\rho_\psi$ term compete with
the $\partial_\phi{\cal U}$ term in the equation of motion for
$\phi$.

Having indicated two types of examples, let us return to the general
analysis of the dynamics.  Let us suppose that enough density
develops that the forces from $V_{eff}/b^4$ dominate over those of
${\cal U}$ for $r$ less than some crossover scale $r_c$.  Let us
start by assuming that the forces on the scalar field are dominated
by field theoretic ones, i.e. that the energy densities in the two
vacua are small enough that the curvature is a negligible
contribution to the scalar equation of motion.  We will also make
use of the discretuum of vacua to assume we can tune the potential
as desired to simplify our analysis.  We will take the local
effective potential $V_{eff}$ for simplicity to have a single
minimum at $\phi=\phi_*$, and ${\cal U}$ to have a minimum at
$\phi=\phi_+$ and a lower minimum at $\phi=\phi_-$. In bulk, $\phi$
is metastabilized at either $\phi_\infty=\phi_+$ or
$\phi_\infty=\phi_-$.


\ifig\casesII{Three basic cases in which a compact object locally
distorts the effective moduli potential.}
{\epsfxsize3.5in\epsfbox{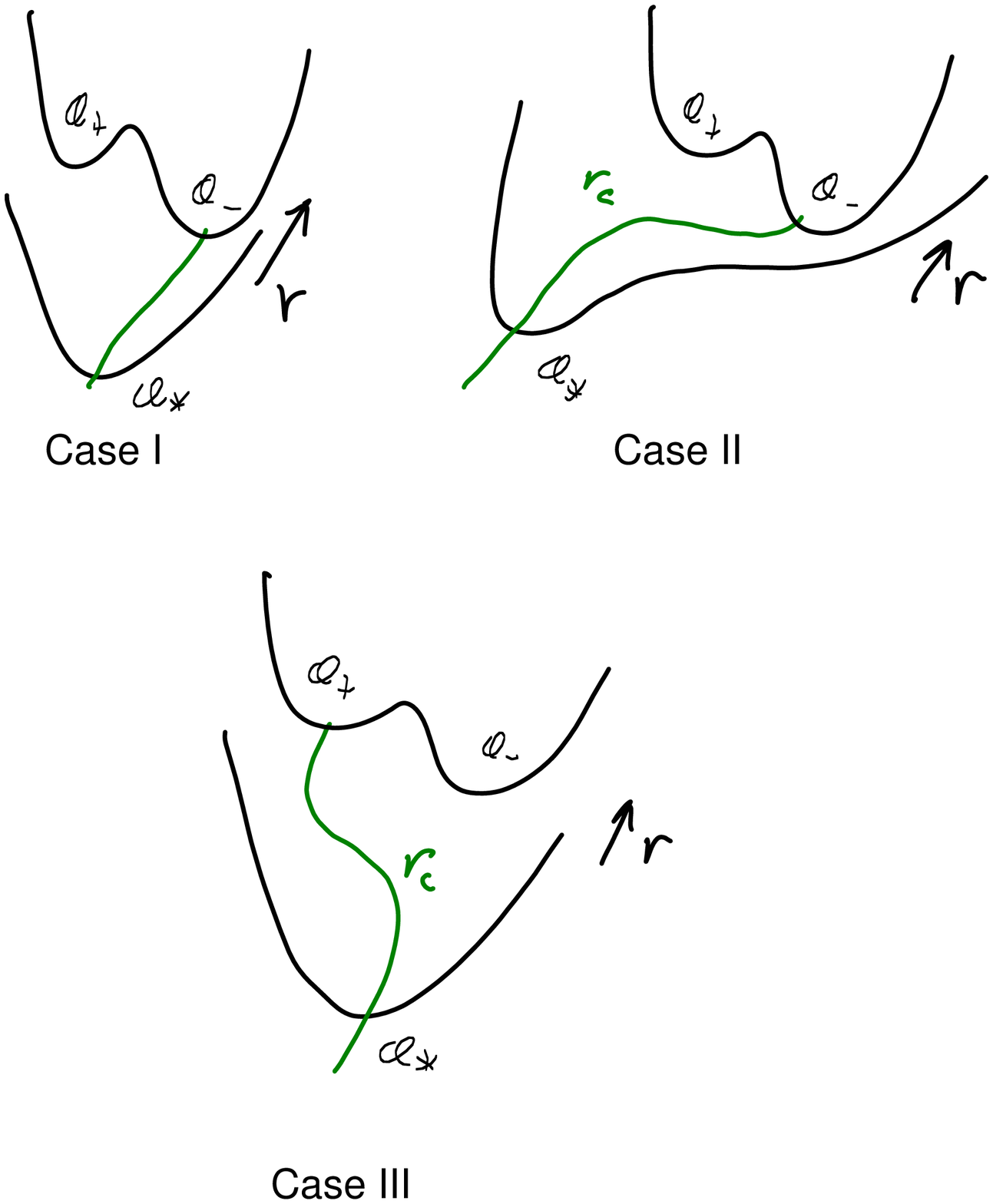}}

There are three basic cases we will discuss (see figure 1): (I)
$\phi_*$ is in the basin of attraction of $\phi_\infty$, (II)
$\phi_\infty=\phi_-$ is in the true vacuum while $\phi_*$ is close
to $\phi_+$, (III) $\phi_\infty=\phi_+$ is in the false vacuum while
$\phi_*$ is close to $\phi_-$.  In cases II and III, one forms a
vacuum bubble.  This bubble is perturbatively stable at a finite
radius in case II and can expand forever for appropriate values of
the parameters in case III.

\subsec{Case I}

First we consider the case where $V_{eff}$ has settled down to a
time independent shape, whose minimum is in the basin of attraction
of ${\cal U}(\phi(r \rightarrow \infty))$, the minimum of ${\cal U}$
where $\phi$ is stabilized asymptotically. There is a static
perturbatively stable solution in this case.  At any given $r$,
there exists a minimum of the combined potential ${\cal
U}_{tot}(r,\phi) = \frac{V_{eff}}{2b(r)^4}+\frac{1}{4}{\cal U}$ that
is continuously connected to minima of ${\cal U}$ and $V_{eff}$.

To first approximation, $\phi$ will sit at the minimum of ${\cal U}_{tot}$ at each $r$:
\eqn\pertphi{ \phi = \phi_{0}(r) + \delta \phi(x) }
where the background solution $\phi_0(r)$ satisfies
\eqn\minform{{\delta{\cal
U}\over{\delta\phi}}\bigg|_{\phi=\phi_0(r)}\equiv 0}
Fluctuations $\delta\phi(x)$ about the minimum are massive.

\subsec{Case II:  perturbatively stable vacuum bubbles}

When $V_{eff}$ has a minimum at the false vacuum value of $\phi$, but $\phi$ is in the true vacuum
asymptotically, a static bubble is formed.

There are two characteristic values of $r$ in this case. As before
$r_c$ is the coordinate radius below which the local effective
potential $V_{eff}/b^4$ draws the scalar field toward $\phi_*$.  A
second radial scale of interest, $r^\prime_c>r_c$ appears in this
case; it is the radial distance at which the two minima of ${\cal
U}_{tot}$ are degenerate:  ${\cal U}_{tot} (\phi_{+}, r^\prime_c) =
{\cal U}_{tot} (\phi_{-}, r^\prime_c)$.

For $r<r_c$, $\phi$ will roll toward the only minimum available.  In
particular, the scalar field VEV will be close to that corresponding
to the false vacuum in bulk. Conversely, if this bubble of false
vacuum extended out to $r \gg r^\prime_c$, then it would
collapse for energetic reasons. Thus, the bubble wall should have a
stable position somewhere near $r^\prime_c$.

The technology developed in \DVB\ could be used to analyze this
case in more detail; we will illustrate this in the next section on case III.

\subsec{Case III:  exploding attractors and vacuum decays}

The final case is where the field is metastabilized in the false
vacuum at infinity, but $V_{eff}$ is in the true vacuum. The
solutions will typically evolve in time, and thus the question of
initial conditions is important. Unlike the previous case, there is
only one characteristic scale $r_c$.

We expect that for $r<r_c$, the field will roll into the minimum.
Therefore, we will start with the bubble wall at a radius slightly
below $r_c$ and we will consider whether or not the bubble expands.
At $r \gg r_c$, ${\cal U}$ dominates and the solution will look like
a typical vacuum bubble so the standard analysis applies to
determine whether the bubble expands. Therefore, we will focus on
the new effects of the local source.  We will be interested in
finding reasonable conditions under which the wall expands from its
starting configuration at $r_c$.

Let us start in pure field theory to gain intuition, after which we
will indicate the fully relativistic generalization. First let us
arrange parameters such that the bubble wall is thin. This means
that $\phi$ varies from $\phi_\infty$ to $\phi_*$ over a radial
distance $\Delta r \ll r_c$.   We follow \DVB\ and work in Gaussian
normal coordinates
\eqn\Gausscoords{ds^2=\epsilon dn^2+\gamma_{ij}dx^idx^j}
where $n$ is the normal direction to the bubble, and $x^i$ the
coordinates along the bubble.  The scalar equation of motion is then
\eqn\scalareom{{1\over 2}{d\over
d\phi}\biggl({d\phi\over{dn}}\biggr)^2+\Gamma_{in}^i{d\phi\over{dn}}={\partial
{\cal U}_{tot}\over{\partial\phi}}}
A thin wall requires that $\biggl({d\phi\over{dn}}\biggr)^2\gg
\Gamma_{in}^i{d\phi\over{dn}}$ at the wall.  This integrates to
\eqn\solphi{\biggl({d\phi\over{dn}}\biggr)^2=2{\cal
U}_{tot}(\phi;\tau)+C}
where $\tau$ is the proper time along the wall and $C$ is an
integration constant.  Setting the variation of $\phi$ with respect
to $n$ to be zero inside the bubble, we can take $C$ to be $-2{\cal
U}_{tot}(\phi_*)$.

Suppose the bubble starts at rest at radius just below $r_c$,
forming once the density accumulates to the point that $V_{eff}/b^4$
dominates over ${\cal U}$ for $r<r_c$.  Let us expand ${\cal
U}_{tot}(\phi;r_c)=V_0+V_1\phi+\dots$, with $V_1$ parameterizing the
steepness of the potential which drives the field toward its inner
vacuum.  To begin with, the normal direction is the $r$ direction,
and the equations integrate to
\eqn\localphi{\phi-\phi_*={V_1\over 2}(r-r_c)^2}

If we fix the difference in field VEVs $\Delta\phi\equiv
\phi_\infty-\phi_*$ and increase the strength of the potential
(increase $V_1$), then we decrease the thickness $\Delta r$ of the
wall and improve the thin wall approximation.

In the example of attractor black holes \Veff, this can be obtained
by choosing sufficiently large charges $Q$ such that $\Delta r \ll
r_c$.  From \scaleq\Veff\ we obtain that the crossover scale $r_c$
occurs at $r_c\sim \sqrt{Q}/M_{\cal U}$ where $M_{\cal U}^4$ is the
scale of the potential ${\cal U}$.  So holding fixed $\Delta\phi$
and scaling up $Q$ increases $r_c$ and decreases the thickness,
improving the thin wall approximation.

Now in order to understand if the bubble will expand, we need to
compare the energy cost from the tension with the energy gained by
replacing the volume inside the bubble by the phase of lower energy
${\cal U}(\phi_*)$.  We will mostly analyze this with a generic
potential characterized by the two scales $M_U, M_0$, in particular
taking the energy difference ${\cal U}(\phi_+)-{\cal U}(\phi_-)\sim
M_U^4$.  In a realistic example with ${\cal U}(\phi_+)$ tuned to be
very small, this generic energy difference applies for decays to a
$\Lambda<0$ phase.  A smaller energy difference corresponds to a
larger critical bubble size required for expansion in the regime
where ${\cal U}$ dominates \CdL.

The tension costs an amount
\eqn\Eten{E_B=4\pi^2 T_B(R_B) R_B^2}
where we have taken into account the fact that the tension will
change with wall radius $R_B$ because our potential ${\cal U}_{tot}$
changes with distance from the source.  In particular, the barrier
between the two basins of attraction $\phi_\pm$ appears for large
enough $R_B$ that ${\cal U}$ begins to dominate in ${\cal U}_{tot}$.

Plugging the solution \localphi\ into the expression for the wall
tension:
\eqn\tension{T_B=\int_{\phi_*}^{\phi_\infty}d\phi\sqrt{2|{\cal
U}_{tot}(\phi)-{\cal U}_{tot}(\phi_\infty)|}}
we obtain at $R_B=r_c$
\eqn\tensionII{T_B(R_B=r_c) = {2\over
3}\sqrt{2V_1}(\Delta\phi)^{3/2}}

Let us first compare the tension energy at $R_B=r_c$ to the
potential energy $E_{bulk}$ liberated by the bubble volume $V_B\sim
r_c^3$. This is
\eqn\Ebulk{E_{bulk}(R_B=r_c) \sim r_c^3 V_1\Delta\phi \gg T_B r_c^2}
which parameterically beats the wall energy $T_B r_c^2$ given our
thin wall parameters $r_c\gg \Delta r$.  More generally, if we stay
outside the compact source (e.g. if we stay outside the horizon of
the black hole), we should consider a shell of bulk potential energy
of thickness $\Delta\tilde r$ with volume $\Delta\tilde r r_c^2$
rather than the whole volume $r_c^3$. The ratio of the bulk energy
to the wall energy is of order
\eqn\compare{{E_{bulk}\over E_{wall}}\sim {\Delta\tilde r\over\Delta r}}

There is also stress-energy from the black hole and from kinetic
energy of the scalar field as it rolls toward the bottom of its
local potential, and from particles that it produces in the process.
Some outgoing radiation will be produced in the process of bubble
formation, as well as ongoing Hawking particle production, leading
to a net flux outward of energy from the compact object.  As we will
discuss momentarily, these effects aid the expansion of the bubble,
which may play a significant role in cases where the bubble would be
sub-critical from the point of view of ${\cal U}$ alone. First,
however, let us continue to assess the expansion based on ${\cal U}$
in the cases where this is sufficient.

Once enough energy has escaped that the black hole horizon is
separated from the wall by a distance $\Delta\tilde r\gg \Delta r$,
the comparison \compare\ means that the wall begins to expand. The
tension of the wall at larger radius increases somewhat due to the
barrier in ${\cal U}$, but as long as the bubble size and the
parameters in ${\cal U}$ are in the range leading to expansion
according to the standard Coleman-de Luccia analysis \CdL, the
explosion will continue unabated.

In the case of \Aac, where the local density arises from massive
particles $\psi$ whose mass depends on $\phi$, the expanding bubble
may produce further $\psi$ particles, amplifying the effect.  This
is because as $\phi$ rolls toward the new minimum $\phi_-$, the mass
of the massive particles decreases.  This can lead to quantum
production of further $\psi$ particles, enhancing the local density
trapping $\phi$ near $\phi_*$ \trapping.

The techniques developed in \DVB\ provide a generally covariant
description of the evolution of bubbles in a way which is convenient
for incorporating the mass and charge of the black hole, and the $r$
dependence in our total effective potential ${\cal
U}_{tot}(\phi;r)$.  The results of this section appear from such an
analysis in the weak gravity limit.

As mentioned above, the inside of the bubble (and the wall itself)
can contain additional sources of energy and pressure.  Thus far, we
have focused on the effective potential for $\phi$ while ignoring
that it originates from a local source with a different equation of
state from the moduli potential.  Moreover, the relaxation of the
$\phi$ field to its local minimum $\phi_*$ proceeds via transfer of
its kinetic energy into particles (including further $\psi$
production as in \trapping).   As such, the pressure inside the wall
has a different relation to the energy density than would arise from
pure vacuum energy. Also, a gauge field confined to the bubble
contributes a positive pressure term that would not appear if it
were simply a scalar. This positive pressure inside the wall helps
the bubble expand. Similar couplings to other standard model
particles will also aid the expansion.

In what follows, we will make several applications of the basic
process we have discussed in this section.  First we will discuss
the stimulated emission of vacuum bubbles in the process of black
hole formation and evaporation.  The charged case \S3.1\ is most
closely analogous to \garybubbles, but we will also find a mechanism
for perturbative production of vacuum bubbles by uncharged bodies
including Schwarzschild black holes \S3.2.  We will then move on in
\S4\ to discuss some implications of this for the landscape, and
make preliminary comments on realistic constraints.

\newsec{Black Holes as Catalytic Vacuum Converters}

\subsec{Vacuum bubbles as endpoints of Hawking decay:  the charged
case}

One application is to generalize the endpoint of Hawking decay found
in \garybubbles. Consider starting in a metastable vacuum at
$\phi_\infty$ and forming a black hole with mass much greater than
its charge. At first, the field $\phi$ is well stabilized at
$\phi_\infty$.  Once the black hole radiates down to the point that
its horizon at $r=r_h$ falls below $r_c$, the local effective
potential $V_{eff}/b^4$ begins to pull the scalar field toward the
point $\phi_*$ minimizing the local effective potential $V_{eff}$.
In case II, this produces a metastable vacuum bubble (a vacuon). In
case III, it produces an explosive vacuum decay. After the vacuum
bubble is produced, the $r_h<r_c$ black hole continues to evaporate,
pair producing particles in the spectrum of the inner vacuum.


Let us next compute the basic thermodynamic quantities at the point
that the bubble is unleashed. This occurs when $r_c\sim
\sqrt{Q}/M_U$ is of order the horizon radius $r_h$. For the
Reissner-Nordstrom case,
\eqn\RN{ds^2=-\biggl(1-{2M\over{M_P^2r}}+{Q^2\over{M_P^2r^2}}\biggr)dt^2+{dr^2\over
\Bigl(1-{2M\over{M_P^2r}}+{Q^2\over{M_P^2r^2}}\Bigr)}+r^2d\Omega^2}
the outer black hole horizon is at $r_h={M\over
M_p^2}+\sqrt{{M^2\over M_P^4}-{{Q^2}\over M_P^2}}$.

When the outer horizon shrinks to $r_c$, the bubble emerges. As in
\garybubbles, this can easily happen far away from the
correspondence point, in the nonextremal regime $M/M_P\gg Q$. In
this regime, the horizon radius is approximately $r_h\sim 2M/M_P^2$.
Setting this equal to $r_c\sim \sqrt{Q}/M_U$ yields $Q\sim
(M^2M_U^2)/M_P^4$. For this to be self-consistent, $Q\ll M/M_P$
implies $M_U^2\ll M_P^3/M\sim M_PM_U/\sqrt{Q}$ or more simply
$M_U\ll M_P/\sqrt{Q}$. Given this, the bubble is released at a point
when the black hole is still very Schwarzschild-like, with
temperature $T\sim M_U/\sqrt{Q}$ and entropy $S\sim QM_P^2/M_U^2\gg
Q^2$.

It would be interesting to obtain a microphysical accounting of
these objects and their explosions, a topic to which we will return
briefly in the discussion section.

There are clearly many variants of this.  For example, the modulus
may couple to other fields which condense inside the bubble,
spontaneously breaking bulk symmetries.  This is a feature of the
tachyon condensation in \garybubbles.  Instead of pure
Reissner-Nordstrom one may consider multiple charges, such as those
combinations for which a standard attractor black hole arises at the
end of the process, surrounded by a vacuum bubble.

\subsec{Schwarzschild black holes and bubbles}

The process \garybubbles\ and its generalization in \S3.1\ depend on
having a charged source for the moduli. The vacuum Schwarzschild
solution does not locally source scalar moduli outside the horizon.
However, the moduli can be sourced by the dense matter coallescing
to form the black hole.  This can happen either before the horizon
forms, or inside the horizon, depending on the energy scales.
Moreover, in the process of Hawking decay the temperature increases
to the point where massive particles get produced; as we will see
these may also source moduli and yield a vacuum bubble.

\subsubsec{Bubble Catalysis in Black Hole formation I:  initial
collapse}

Matter which collapses to form a black hole of mass $M$ develops a
density of order $M/R_S^3\sim M_P^6/M^2$ when the matter reaches the
Schwarzschild radius $R_S\sim M/M_P^2$.  If this density competes
with the barrier heights in the moduli potential $\cal U$, and if
the local and ambient potentials are configured as in case II or III
above, then this process will produce a vacuum bubble before a black
hole is formed.  In case II, it creates a perturbatively stable
vacuon, and in case III an explosive vacuum decay.

\subsubsec{Bubble Formation in Black Hole formation II:  inside}

In many models, the black holes which form by structure formation
have a Schwarzschild density $M_P^6/M^2$ which is very low compared
to natural barrier heights.  For example, the smallest black holes
inferred from core collapse events have roughly solar mass,
corresponding to an energy density of $GeV^4$. This means that for
${\cal U}\gg GeV^4$, no bubbles are formed outside the horizon in
their formation.

However, inside the horizon, bubbles are generically classically
nucleated by the crunching matter source. In particular, in weakly
coupled string landscape models \land, there is a runaway direction
for the dilaton and/or volume moduli separated by a barrier from
metastable vacua. These moduli will generically couple to the masses
of particles forming the black hole.  As a result, a vacuum bubble
containing the basin of attraction of the large volume/weak coupling
limit will materialize inside the black hole.  The Kasner solution
approaching the black hole singularity must therefore be extended to
include the rolling dilaton, volume, and other liberated moduli of
the compactification inside the bubble.\foot{Note that this result
is different from the suggestion \smolin\ that new universes are
created inside black holes, which would require the black hole
singularity to be resolved in such a way as to connect a big crunch
with a big bang.  Evidence such as
\refs{\garyjoe,\spacelike,\finalstate}\ contraindicates the
hypothesis of a crunch-bang inside black holes.  In any case, our
results about bubble formation {\it outside} black holes support the
idea that black holes induce mixing between different metastable
vacua; our mechanism for producing such mixing is the much more
prosaic classical catalysis effect described in the text.}

\subsubsec{Bubble formation in Black Hole Evaporation III:  massive
Hawking particles seed bubbles}

Let us consider a model such as \Aac\ and simple generalizations in
which $\phi$ modulates the mass of a set of fields such as $\psi$.
In the process of evaporation, the black hole eventually reaches
high enough temperatures $T$ to produce massive Hawking particles
$\psi$ of mass $m_\psi$. Once $T$ reaches the threshold $T \sim
m_\psi$, a significant density of the massive $\psi$ particles is
produced \page.  If these particles do not decay or disperse too
rapidly, they form a density of particles which can kick $\phi$ from
one basin of attraction to another.  We will now assess a window of
parameters where this occurs.  For simplicity we will first focus on
a regime where the produced massive $\psi$ particles are still
non-relativistic far from the BH (with outward velocities beating
the escape velocity, but much smaller than the speed of light).  The
black hole also produces quanta of $\phi$ itself. For consistency we
will then check that $\phi$ fluctuations do not themselves produce
bubbles, and that they do not wash out the $\psi$-catalyzed bubble
production in the range of parameters of interest. With $m_\psi\gg
m_\phi\sim M_U^2/M_0$, the situation is schematically depicted in
figure 2.

\ifig\cloudsII{In a time period $\Delta t$, the black hole (filled
circle) produces a gas of massive $\psi$ particles which spreads a
distance $\Delta r$ to occupy the region indicated.  For the regime
of parameters discussed in the text, this seeds a vacuum bubble.
During the same period, the lighter $\phi$ fluctuations disperse
more rapidly and have diluted to occupy a smaller energy density
over a larger region.  For an appropriate window of parameters the
fluctuations of light $\phi$ particles do not wash out the bubble
nucleated by the gas of massive $\psi$ particles.}
{\epsfxsize3.5in\epsfbox{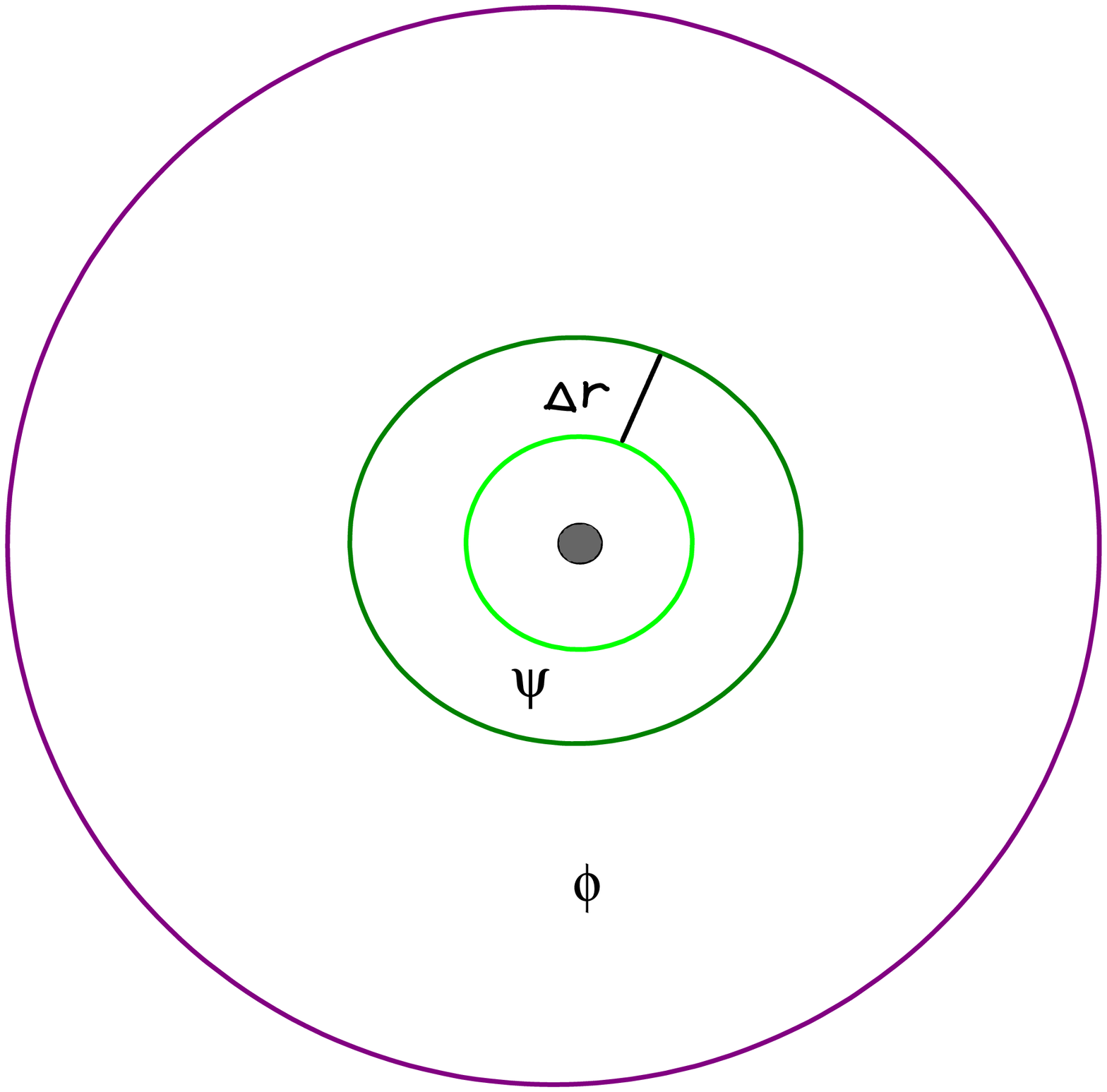}}


There are five relevant scales in the problem:  $M_P$, $M_0\sim
|\phi_+-\phi_-|$ (the barrier width in field space), $\tilde M_0$
(determining the $\psi-\phi$ coupling in \Aac), $m_\psi$ (which will
be of order the temperature $T$ in the regime of interest), and
$M_U$ (the energy scale in the potential ${\cal U}\sim M_U^4
f(\phi/M_0)$).

We will consider moduli for which $M_U\ll M_0$. This yields weak
self-interactions in ${\cal U}$.  The couplings between $\psi$ and
$\phi$ go like $m_\psi/\tilde M_0$ \Aac. For the catalysis effect we
will be led to a window in which $m_\psi\ge M_0$, requiring us to
take $\tilde M_0 > M_0$ to avoid $\ge {\cal O}(1)$ interactions
between $\psi$ and $\phi$.  In order to avoid large renormalizations
of ${\cal U}$ due to these couplings, one can also consider a
supersymmetrized version of the model with a SUSY breaking scale
much lower than $M_0$.

The energy density in $\psi$ particles is of order
\eqn\endensI{\rho_\psi\sim {{\Delta M\xi}\over (Vol)}}
where $Vol$ is the volume of space into which the produced $\psi$
particles have dispersed in a time window for which the black hole
mass has decreased by $\Delta M$, of which a fraction $\xi$ has gone
into the production of $\psi$ particles.  In this time window
$\Delta t$, the particles spread out a distance $\Delta r$, which we
will insist be much greater than the black hole size $T^{-1}$. This
ensures that the particles we consider are far from the black hole
horizon, so that the Hawking calculation of the asymptotic produced
particle distribution is accurate.  More specifically, let us count
the particles emerging from the black hole starting at a distance
$\Delta r\gg 1/T$, and consider their further spread into the volume
contained between $r=\Delta r$ and $r=2\Delta r$.

We will insist that $\Delta r$ be at least as large as the critical
bubble size $R_c\sim M_0/M_U^2$ above which the bubble expands in
the regime where the potential ${\cal U}$ dominates (as derived in
the standard analysis \CdL\ and reviewed in \S2). Given $\Delta r\gg
T^{-1}$, the volume $Vol$ in
\endensI\ is of order $\Delta r^3$.  We will find conditions under
which the density \endensI\ is competitive with (or stronger than)
the ambient potential ${\cal U}$, and that the time window $\Delta
t$ is sufficiently long that the local density
\endensI\ can drive $\phi$ across the barrier, a distance $M_0$ in
field space.

As the black hole temperature $T$ increases to the threshold $T =
m_\psi$ to produce $\psi$'s, at first it produces them
non-relativistically. Their velocity is then of order
\eqn\nonrelv{v=\sqrt{2\biggl( {E_\psi\over m_\psi}-1 \biggr)}}
using ${1\over 2}m_\psi v^2=E_\psi-m_\psi$. As long as the massive
particles are non-relativistic, their velocity remains small and
their dispersal is correspondingly slow allowing them to form a
dense source for $\phi$ (as we will see in what follows). The
non-relativistic approximation remains valid as long as we consider
a period in which $T$ remains close enough to $m_\psi$ that
$E_\psi\sim T$ satisfies
\eqn\Tbound{{{E_\psi-m_\psi}\over m_\psi}\equiv {\Delta T\over T}\ll
1}

From this we can obtain the fraction $\xi$ of the black hole
emission which is contained in non-relativistic $\psi$ particles, as
follows. Hawking evaporation produces a thermal distribution of
particle numbers \page, up to the greybody factor arising from the
absorption cross section $\sigma(E)$.  For a given species, the
number $N$ of particles produced by the black hole as a function of
time is given by
\eqn\emission{{dN\over{dt}}={v \sigma(E)\over{e^{E/T}\pm 1}}{d^3\vec
k\over{(2\pi)^3}}={\sigma(E)\over{e^{E/T}\pm 1}}{v^2
E^2dE\over{2\pi^2}}}
where $\sigma(E)$ is of order $1/(T^2v^2)$ \crosssection.

For a nonrelativistic species such as our $\psi$ particles,
integrating this from $E=m_\psi$ to $E\sim T=m_\psi+{1\over 2}m_\psi
v^2$ yields $dN/dt$ of order $(v \sigma (E)) v^3 m_\psi^3\sim v^2
m_\psi^3/T^2$. For a relativistic species, integrating it from $E=0$
to $E\sim T$ gives a result for $dN/dt$ of order $\sigma (E)
m_\psi^3\sim m_\psi^3/T^2 $.

So the non-relativistic emission of $\psi$ particles is down by a
factor of order
\eqn\efficiency{\xi\sim v^2\sim {\Delta T\over T}}
from the total emission of the black hole in this window.

To obtain the other factors in  $\rho_\psi$ \endensI, we must relate
$\Delta M$ and $\Delta r\sim (Vol)^{1/3}$ to $\Delta T/T$.  This
goes as follows. The change in mass is
\eqn\masschange{\Delta M\sim M_p^2\biggl({1\over
m_\psi}-{1\over{m_\psi+\Delta T}}\biggr)\sim {M_p^2\over
T}\biggl({\Delta T\over T}\biggr)}
As discussed above, we will consider a window such that $\Delta r$
is at least as big as the critical bubble size $R_c\sim M_0/M_U^2$
in the potential ${\cal U}$.  So set
\eqn\nextrelns{\Delta r=\eta M_0/M_U^2 \sim v\Delta t \sim \Delta
t\sqrt{\Delta T/T}}
with $\eta\ge 1$, and relate $\Delta t$ to $\Delta T$ as follows.
The window of temperatures in the range $(m_\psi, m_\psi+\Delta T)$
corresponds to the time period
\eqn\trange{\Delta t \sim {M_p^2\over 3}\biggl({1\over
T^3}-{1\over{(T+\Delta T)^3}}\biggr)\sim \biggl({\Delta T\over
T}\biggr){M_p^2\over T^3}}
obtained by integrating the Stefan-Boltzmann law $dM/dt\sim
(area)\times T^4 \sim T^2$ using $M\sim M_P^2/T$.

Plugging \trange\ into \nextrelns, we obtain
\eqn\TrangeII{v^3\sim \biggl({\Delta T\over T}\biggr)^{3/2}\sim
{{T^3M_0\eta}\over{M_P^2M_U^2}}\ll 1}
with the last inequality enforced for self-consistency of the
non-relativistic approximation.

Putting these estimates together, we obtain
\eqn\endensresult{\rho_\psi\sim M_U^4{v T^2\over {\eta^2 M_0^2}}}
We will shortly impose the condition that this density not decay
rapidly by annihilation into $\phi$ particles.  First let us proceed
to analyze its effects given this.

In order for the forces on $\phi$ from this energy density
$\rho_\psi$ to compete with the ambient potential ${\cal U}\sim
M_U^4 f(\phi/M_0)$, we require $(\del_\phi\rho_\psi\sim
\rho_\psi/\tilde M_0)
> (\del_\phi {\cal U}\sim M_U^4/M_0)$ which translates into the statement
\eqn\Tineq{v T^2 \ge \eta^2 M_0\tilde M_0}
This is consistent with our condition for perturbativity, $\tilde
M_0\ge (T\sim m_\psi)\ge M_0$.

From \Aac\ we see that the dimensionless couplings between $\psi$
and $\phi$ are of order $m_\psi/\tilde M_0$.   Since $T\sim m_\psi$,
if we took $\tilde M_0\sim M_0$, \Tineq\ would lead to $\ge {\cal
O}(1)$ couplings between $\phi$ and $\psi$.  In order to avoid that,
we may specify that $\phi$ couples to $\psi$ via weaker (e.g.
$\tilde M_0\sim M_P$-suppressed) couplings than appear in ${\cal
U}$.

It is also important to check that the mean field treatment of the
$\psi$ energy density is appropriate.  That is, we must check that
the spacing $L$ between $\psi$ particles is less than the Compton
wavelength $m_\phi^{-1}\sim M_0/M_U^2$ of $\phi$.  For the minimal
density that competes with the moduli potential, number density of
$\psi$ particles is of order
\eqn\numdens{n_\psi\sim {\rho_\psi\over m_\psi}\sim {\rho_\psi\over
T}\sim {{M_U^4\tilde M_0}\over{TM_0}}\sim {1\over L^3}}
Setting $1/L^3\gg m_\phi^3$ translates into the requirement
\eqn\comptfinal{(M_U/M_0)^2(T/\tilde M_0)\ll 1.}

We must also check that the time window $\Delta t$ \trange\ is
sufficiently long that the density $\rho_\psi$ has time to kick
$\phi$ across the barrier.  We can relate $\Delta t$ to the range of
field space $\Delta\phi$ that $\phi$ rolls during the process as
follows.  As before let us expand the potential ${\cal U}_{tot}\sim
V_0+V_1\phi+\dots$.  In a local region, the scalar rolls according
to $\del_t^2\phi\sim V_1$ \ie\
\eqn\phirange{\Delta\phi\sim V_1(\Delta t)^2}
In the case that the accumulated $\psi$ density is of order $M_U^4$
(and hence competitive with the ambient moduli potential), we can
identify $V_1\ge M_U^4/M_0$ yielding
\eqn\phirangeI{\Delta\phi\ge {M_U^4\over M_0}(\Delta t)^2\sim
M_0\eta^2 {T\over\Delta T}.}
which is automatically greater than $M_0$ in our non-relativistic
regime.  So the window $\Delta T$ is easily long enough to drive
$\phi$ into the basin of attraction of a different minimum.

Now let us address the decay of the density \endensresult\ via
annihilation of $\psi$ particles into $\phi$ particles.  As we noted
above, a natural model in which to apply this would be a low energy
supersymmetric model, for which $\psi$ has scalar superpartners also
with mass of order $m_\psi$. The $2\to 2$ center of mass scattering
cross section $\sigma$ for scalars scales like
\eqn\crosssec{\sigma\sim \lambda^2 {k_f\over k_i}{1\over 4\pi E^2}}
where $\lambda\sim {m_\psi^2\over \tilde M_0^2} $ is the quartic
coupling, $k_f$ is the final momentum (of order $m_\psi$ in our
problem), and $k_i$ is the initial momentum scale (of order
$vm_\psi$ in our case). Let us choose $\tilde M_0$ large enough that
this is $\le 1$, i.e. $\tilde M_0\ge m_\psi$.

Given the $\psi$ density $n_\psi\sim \rho_\psi/m_\psi$
\endensresult, velocity $v$ \nonrelv\Tbound, and cross section
$\sigma$ \crosssec, the annihilation rate for a given $\psi$
particle is
\eqn\nvsigma{\Gamma = n_\psi v \sigma }
Setting
\eqn\anncond{\Gamma \Delta t \ll 1}
ensures that the $\psi$ particles do not typically annihilate during
our window of interest.  In terms of our parameters, this is the
condition
\eqn\anncondII{{{v^3M_P^2M_U^4}\over{\eta^2\tilde M_0^4M_0^2}}\ll 1}
The fermion annihilation cross section is down from the scalar one
\crosssec\ by a factor of $p_i/m_\psi\sim v \ll 1$.

Finally let us check the effects of $\phi$ fluctuations themselves.
First, let us note that the gases of particles produced by the black
hole (including both $\phi$ fluctuations and the $\psi$ particles in
our model) are not in thermal equilibrium at temperature $T$ in the
regime of interest.  The particles are produced by the hot black
hole, but spread out to radii of order $\Delta r\gg 1/T$ and hence
their energy density is not that of an equilibrium thermal gas at
temperature $T$.

We want to check if the $\phi$ particles produce a substantial back
reaction on the system. We will address the effects of the $\phi$
fluctuations in both the non-relativistic and relativistic regimes.

Expanding our potential ${\cal U}\sim M_U^4  f(\phi/M_0)$ we have a
$\phi$ mass $m_\phi\sim M_U^2/M_0$. As the black hole heats up past
the threshold $T_\phi\sim M_U^2/M_0$ for producing $\phi$
perturbations, it begins to do so non-relativistically similarly to
our discussion above. However there we saw \Tineq\ that $T\gg M_0$
was required to catalyze a bubble of radius greater than the
critical size (assuming the effective potential in the presence of
the gas of particles drives $\phi$ in this direction). So for
$M_U<M_0$, as happens for weakly coupled moduli, the regime where
non-relativistic $\phi$ particles are produced does not in itself
produce a $\Delta r$-sized bubble.

In the regime $T\gg m_\phi $ in which the $\phi$ fluctuations are
produced relativistically, the typical momentum of produced $\phi$
particles is of order $k\sim T\gg M_U^2/M_0$.  The kinetic energy
$k^2\phi^2$ is much greater than the potential energy
$m_\phi^2\phi^2$.  Similarly to the discussion above, we can
estimate the energy density $\rho_{\delta\phi}$ in relativistic
$\phi$ particles at a radial distance of order the critical radius
$R_c\sim M_0/M_U^2$.  This yields $\rho_{\delta\phi}\sim
M_U^4(T/M_0)^2$. The force on the field comes only from the
potential energy component of this, which is of order
$\rho_{\delta\phi}(m_\phi^2/k^2)\sim M_U^4(M_U/M_0)^4\ll M_U^4$.  So
the effective force in the presence of the $\phi$ particles in the
relativistic regime is also too weak to kick the field across the
potential barrier.

As pointed out in \tometal, the suppression factor $m_\psi^2/k^2$ in
the above estimate would be weakened if the energy scale $k$ were
much lower than the black hole temperature $T$.  This may happen via
thermalization in the presence of sufficiently strong interactions
among the $\delta\phi$ particles. But in the weakly coupled regime
$M_U\ll M_0$ we consider here we find this effect is not
significant.

This result is perhaps not surprising, given that it has been argued
 that symmetry is not restored in the thermal bath seen
 by a non-inertial observer \unruh.  The usual explanation
 for this result is that the acceleration and the temperature
 scale in the same way, so the effective gravity spoils the symmetry
 restoration that occurs in flat space.  What makes the transition possible
 in the case of the $\psi$ particles is precisely that a significant density
 is formed at distances large enough to ignore gravitational effects.  The fact
 that we cannot achieve this with the weakly interacting
 $\phi$ particles may not be a coincidence.

So far we have seen that $\phi$ fluctuations do not themselves
produce large vacuum bubbles; this means the effect we are
considering does require the extra $\psi$ particles. Next let us
check whether the kinetic-energy-dominated spatially varying $\phi$
fluctuations wash out the bubble produced by the $\psi$ gas. The
average field fluctuation $\delta\phi$ is determined by
$k^2\delta\phi^2\sim\rho_{\delta\phi}$, with $k\sim T$.  This could
potentially be a problem if the distance in field space $\delta\phi
\sim \sqrt{\rho_{\delta\phi}}/T$ is greater than or equal to the
barrier width $M_0$.   In the same window considered above for
$\psi$-catalyzed bubble production, the density in $\phi$ particles
is of order
\eqn\endensphi{\rho_{\delta\phi}\sim M_U^4 {T^2\over \eta^2 M_0^2}}
Setting this less than $M_0^2T^2$ ensures that $\delta\phi<M_0$ so
that the $\phi$ fluctuations do not wash out the $\psi$ catalysis
effect.  This is automatic for $M_U<M_0$.

In the regime of parameters we have taken the $\psi$ catalysis
occurs classically once the $\psi$ particles have been created, so
this effect automatically dominates over exponentially suppressed
thermal and tunneling effects.

It is readily verified that a window of parameters exists where all
the above constraints are satisfied.  As a specific example, the
following hierarchy of scales works:
\eqn\achoice{\eta=10 ~~~~ {M_0\over M_U}\sim 10^2 ~~~~ {m_\psi\over
M_U}\sim 10^8 ~~~~ {\tilde M_0\over M_U}\sim 10^9 ~~~~ {M_P\over
M_U}\sim 10^{16}}

Altogether these estimates suggest that vacuum bubbles can also
appear as new endpoints of Hawking radiation in the case of
uncharged black holes! Whatever formed the black hole, the process
of evaporation proceeds through higher and higher temperatures,
eventually producing massive particles which can seed a vacuum
bubble surrounding the evaporating black hole if the mass scales lie
in the range satisfying the above constraints.

The above estimates are in fact somewhat more conservative than
necessary.   For example, it is possible to extend this mechanism to
the regime where the produced $\psi$ particles are relativistic. It
is only the $m_\psi\bar\psi\psi$ contribution to the stress energy
which sources the field, which is down by a factor of $(m_\psi/T)^2$
from the full energy density in the relativistic regime.  But the
production of relativistic $\psi$ particles is unsuppressed.
Altogether, requiring the force at $r\sim \eta R_c $ to be greater
than that from ${\cal U}$ leads to a somewhat wider window of
parameters where the effect occurs. For example, the constraint
\Tineq\ becomes
\eqn\relTineq{m_\psi^2\ge \eta^2 M_0\tilde M_0}
Moreover once the bubble is produced (either relativistically or
non-relativistically), further $\psi$ particles -- produced from the
black hole and from the rolling $\phi$ field -- push it out further.

\newsec{Other applications}

\subsec{Population and stability of the landscape}

We have just found a significant range of parameters for which
Schwarzschild black holes ultimately catalyze vacuum bubbles in the
process of Hawking evaporation if not before, since they produce
massive particle densities that source the dilaton runaway
direction.  The decay of large black holes, while a very long
process, is parameterically faster than bubble nucleation by
tunneling.

This effect must be taken into account in assessing stability of
metastable vacua.  Realistic application requires tuning the
cosmological constant to be small, and then comparing the particle
spectrum and parameters in the potential to those required above for
catalyzed decay. Transitions from a realistically small cosmological
constant to a $\Lambda=0$ minimum involve a very small bulk
potential energy difference, which makes it more difficult to obtain
an explosive decay. However transitions from realistic cosmological
constant vacua to nearby $\Lambda<0$ phases are not so suppressed,
and the above mechanism can destabilize the model well before
tunneling events do.

In the context of the landscape, this appears to be the dominant
instability in a significant range of backgrounds.  For example, in
the case of our universe with solar mass black holes formed from
core collapse events, the decay time arising from the evaporation of
these black holes, of order $\sim 10^{65}$ years, is substantially
shorter than the decay time of order $e^{10^{120}}$ obtained from
tunneling in appropriate regions of parameter space \land\ (though
still safely longer than the age of the universe). This alone, while
leading to a vastly shorter decay timescale, does not constrain the
models realistically.  However in the context of models leading to
denser objects \refs{\Qball,\graveyard}, there may be
phenomenologically significant constraints from this process \Qball.

One can also apply this to the early universe as a mechanism for
populating the landscape. Backgrounds in which overdensities develop
which compete with the barriers in ${\cal U}$ will experience
catalyzed vacuum bubble production. The formation of dense
structures is a somewhat delicate process; in the observable
universe the formation of dense stars inside gravitational potential
wells induced by dark matter depends on appropriate cooling
mechanisms (which may then collect yet denser structures
\refs{\Qball,\graveyard}).  In any case, in the apparently vast
discretuum of string vacua, there may be many examples with dense
structures forming in hidden sectors, which lead to vacuum bubble
production. In the case (III) of explosive vacuum decay, this
process populates the landscape faster than occurs via tunneling.

This produces a dynamical trend toward solutions with smaller
inhomogeneities: backgrounds with large inhomogeneities seed vacuum
decays, producing new backgrounds which produce their own bubbles
until the process shuts off with the production of backgrounds with
small inhomogeneities relative to barrier heights.

This dynamics also produces bubbles with lower mass particles, since
the forces that draw $\phi$ toward the new vacuum at $\phi_* $ arise
by virtue of particles whose mass decreases as $\phi\to\phi_* $.
Although our mechanism here is different, the same trend as in
\trapping\ toward points with extra light particles arises in this
context. Moreover, the bubble need not be empty; as the bubble
expands and the field rolls toward $\phi_*$, the masses of the
$\psi$ particles change with time and they may get produced.

\subsec{Vacuum bubble production in realistic models: constraints
and corners}

In a highly model-dependent way, vacuum bubbles may be produced by
catalysis in realistic scenarios.  This depends on the densest
structures that form in the model.

Some inflation models such as certain hybrid inflation models
\andrei\ produce small primordial black holes.  In the context of
the landscape, their formation and decays can produce vacuum bubbles
instead of simply leaving behind small bursts of radiation.  For
case III this constrains inflationary parameter space to some
extent, and conversely in case II it provides a model-dependent
mechanism for producing vacuum bubbles.

A more interesting regime where dense structures may form \Qball\ is
via the collection and coalescence of charged exotics in stars,
generalizing the effect discussed in \graveyard.

Another natural question is whether real world core collapse events
can produce transient vacuum bubbles; in general it is of interest
to explain type II supernova explosions \woosley.\foot{Heterotic
supernovae have yet to be detected.} Scalar field moduli typically
couple to standard model fields. These may be stabilized at a very
high scale, suppressing catalyzed decays. However, very low
potentials can be obtained technically naturally, for example in the
case that a scalar field $\phi$ couples preferentially to the
neutrino mass term (cf \mavens) in a similar way at low energies to
the example \Aac. In this case, one obtains a potential energy of
order $m_\nu^2M_C^2$ where $M_C$ is the effective cutoff in neutrino
loop contributions to the moduli potential.  If the latter is at the
supersymmetry breaking scale of $TeV$, the resulting energy scale
for the radiatively generated moduli potential is roughly of order
$MeV^4$, \ie\ $M_U\sim MeV$.

The densest known structures are the neutron stars formed from core
collapse events.  The energy density in these environments is of
order the QCD scale; the neutrinos form a trapped degenerate
relativistic gas with chemical potential of order 200 MeV and hence
energy density $(200 MeV)^4$ in the core \raffelt.  However, the
scalar fields modulating the neutrino masses are directly sensitive
only to the nonrelativistic correction to the energy density.  This
is down by a factor of order $m_\nu^2/p^2\sim 10^{-18}$ for momentum
scale $p\sim 200 MeV$ from the relativistic energy $(200 MeV)^4$.

Hence absent further cancellations, the mass-dependent contribution
to the energy density (to which the scalar is directly sensitive) is
small compared to the natural scale of barrier heights.  However, in
models with low barrier heights, attractor explosions might play a
role.\foot{The present simulations aiming to explain type II
supernovae via neutrino energy deposition do not consistently yield
explosions \woosley, so it is conceivable that new physics will be
required. However it is entirely possible that explosions will arise
from more conventional mechanisms (see e.g. \burrows), so this
rather tuned regime of parameters does not appear well motivated
unless serious puzzles with generating SNe explosions persist.}

\newsec{Discussion}

In this note, we have seen how starting in a single metastable
vacuum, one can assemble compact objects which unveil bubbles of
other vacua. We found large classes of examples generalizing
\garybubbles\ to provide vacuum bubbles as endpoints of Hawking
evaporation of charged and uncharged black holes.  We have also
noted that matter sources can produce bubbles of other vacua inside
the horizon, which modify the internal solution before impinging on
the singularity; their existence must be imprinted in the decay
products of the black hole.

Perhaps most interestingly, we found that in the process of
evaporation of Schwarzschild black holes, the massive particles
produced in the Hawking process can seed vacuum decay.  The black
hole catalytically converts whatever formed it into all the
particles of the system, including massive particles sourcing
moduli.  The resulting local potential forces the moduli into a
different vacuum for a range of model parameters.  This perturbative
effect occurs much more rapidly than exponentially suppressed
thermal and tunneling effects.

Our analysis has been semiclassical, combining structure formation
and gravitational collapse with Hawking radiation and classical
bubble nucleation.  The basic laws of black hole mechanics \BHmech\
are classical, but provided clues pointing toward a more
microphysical statistical description of black holes.  It would be
interesting to determine a dual microphysical description of these
processes.

In general, our main lesson is simple.  Compact objects made in one
vacuum contain information about other vacua, if the objects are
sufficiently dense. This regime seems potentially more accessible
than non-perturbative cosmological methods for connecting present
physics with the other vacua in the landscape. Moreover dual
descriptions of black hole entropy and dynamics must account for
attractor explosions in the generic setting \land. Finally,
assessing the ultimate decay modes of physical models built on
metastable vacua requires analyzing the structure formation in the
models and estimating the catalyzed decay time as well as the
ambient tunneling amplitude.

\bigskip
\bigskip

\centerline{\bf{Acknowledgements}}

We would like to thank T. Banks, M. Douglas, G. Horowitz, S. Kachru,
R. Kallosh, A. Linde, J. Polchinski, S. Shenker, and J. Wacker for
useful discussions and N. Arkani-Hamed, S. Dimopoulos, P. Schuster,
and N. Toro additionally for sharing their insights into vacuon and
black hole formation mechanisms \Qball.  We thank P. Schuster for
extensive discussions on the process of \S3.2, including identifying
a missing factor of $v$. We are supported in part by the DOE under
contract DE-AC03-76SF00515 and by the NSF under contract 9870115.

\listrefs

\end